
%
\input harvmac
\def\figflag{I}
\def\psheader#1{}


\font\blackboard=msbm10 \font\blackboards=msbm7
\font\blackboardss=msbm5
\newfam\black
\textfont\black=\blackboard
\scriptfont\black=\blackboards
\scriptscriptfont\black=\blackboardss
\def\blackb#1{{\fam\black\relax#1}}



%
\def\BC{{\blackb C}} 
 \def\BI{{\blackb I}}
\def\BZ{{\blackb Z}}

%
\font\mathbold=cmmib10 \font\mathbolds=cmmib7
\font\mathboldss=cmmib5
\newfam\mbold
\textfont\mbold=\mathbold
\scriptfont\mbold=\mathbolds
\scriptscriptfont\mbold=\mathboldss
\def\bi{\fam\mbold\relax}


\def\tfig#1{Fig.~\the\figno\xdef#1{Fig.~\the\figno}\global\advance\figno by1}
\def\figI{I}
%
\newdimen\tempszb \newdimen\tempszc \newdimen\tempszd \newdimen\tempsze
\ifx\figflag\figI
\input epsf
%
\def\epsfsize#1#2{\expandafter\epsfxsize{
 \tempszb=#1 \tempszd=#2 \tempsze=\epsfxsize
     \tempszc=\tempszb \divide\tempszc\tempszd
     \tempsze=\epsfysize \multiply\tempsze\tempszc
     \multiply\tempszc\tempszd \advance\tempszb-\tempszc
     \tempszc=\epsfysize
     \loop \advance\tempszb\tempszb \divide\tempszc 2
     \ifnum\tempszc>0
        \ifnum\tempszb<\tempszd\else
           \advance\tempszb-\tempszd \advance\tempsze\tempszc \fi
     \repeat
\ifnum\tempsze>\hsize\global\epsfxsize=\hsize\global\epsfysize=0pt\else\fi}}
\epsfverbosetrue
\psheader{fig.pro}       
\fi
%

%
%
%
%

\def\ifigure#1#2#3#4{
\midinsert
\vbox to #4truein{\ifx\figflag\figI
\vfil\centerline{\epsfysize=#4truein\epsfbox{#3}}\fi}
\baselineskip=12pt
\narrower\narrower\noindent{\bf #1:} #2
\endinsert
}
%
%
\def\ifigures#1#2#3#4#5#6#7#8{
\midinsert
\centerline{
\hbox{\vbox{
\divide\hsize by 2
\vbox to #4truein{\ifx\figflag\figI
\vfil\centerline{\epsfysize=#4truein\epsfbox{#3}}\fi}
\baselineskip=12pt
\narrower\narrower\noindent{\bf #1:} #2
}}\qquad
\hbox{\vbox{
\divide\hsize by 2
\vbox to #8truein{\ifx\figflag\figI
\vfil\centerline{\epsfysize=#8truein\epsfbox{#7}}\fi}
\baselineskip=12pt
\narrower\narrower\noindent{\bf #5:} #6
}}}
\endinsert
}



\def\ket#1{| #1 \rangle}         
%

\def\del{\partial}
\def\delb{\overline{\del}} 
\def\fourpt{\hbox{{$\rangle \kern-.25em \langle$}}} 
\def\tree{\hbox{{$\rangle \kern-.5em - \kern-.5em \langle$}}}
\def\ib{{\bar \imath}} 
\def\jb{{\bar \jmath}} %
\def\kb{{\bar k}}
\def\zb{{\bar z}}\def\yb{{\bar y}}
\def\H#1#2{{\rm H}^{#1}(#2)} 

\def\dd{\mskip 1.3mu{\rm d}\mskip .7mu} 
\def\ie{{\it i.e.}}\def\etc{{\it etc.}}

\def\cp#1{{\BC{\rm P}^{#1}}}

\def\Ka{K\"ahler}
\def\sm{$\sigma$-model}
\noblackbox

\Title{\vbox{\hbox{PUPT--1365}\hbox{\tt hepth@xxx/yymmddd}}}
{Notes on N=2 $\sigma$-models$^\star$}

\centerline{Jacques Distler}\smallskip
\centerline{Joseph Henry Laboratories}
\centerline{Princeton University}
\centerline{Princeton, NJ \ 08544 \ USA}
\bigskip\bigskip

\footnote{}{{\parindent=-10pt\par $\star$
\vtop{
\hbox{Email: {\tt distler@puhep1.princeton.edu} .}
\hbox{Research supported by NSF grant PHY90-21984.}
     }     }}
\def\myab{The first lecture is devoted to the
``special geometry" of the moduli space of $c=9$
N=2 superconformal field theories. An important role is played by the extended
chiral algebra which appears in theories with integer $U(1)$ charges.
The second lecture is devoted to the \sm\ approach. The main focus is an
explication of a calculation of Aspinwall and Morrison.}

These lectures, given at the 1992 Trieste Spring School, are devoted to some
selected topics in N=2 \sm s on Calabi-Yau manifolds and the associated
N=2 superconformal field theories. \myab

\Date{12/92}                 
\lref\SpecialGeo{B. de Wit, P. Lauwers and A. van Pr\oe yen, Nucl. Phys. {\bf
B255} (1985) 569.}
\lref\ranspec{S. Cecotti, S. Ferrara and L. Girardello, Int. J. Mod. Phys.
{\bf A4} (1989) 2475\semi
L. Castellani, R. D'Auria and S. Ferrara, Class. Quantum Grav. {\bf 7} (1990)
1767.}
\lref\CandMod{P. Candelas and X. De la Ossa, Nucl. Phys. {\bf B355} (1991)
455.}
\lref\DKL{L. Dixon, V. Kaplunovsky and J. Louis, Nucl. Phys. {\bf B329}
(1990) 27.}
 \lref\StromSpec{A. Strominger, Comm. Math. Phys. {\bf 133}
(1990) 163.}
\lref\PerStrom{V. Periwal and A. Strominger, Phys. Lett. {\bf B335} (1990)
261.}
\lref\TopAntiTop{S. Cecotti and C. Vafa, Nucl. Phys. {\bf B367} (1991) 359.}
\lref\AandB{E. Witten, in ``Proceedings of the Conference on Mirror Symmetry",
MSRI (1991).}
\lref\LVW{W. Lerche, C. Vafa and N. Warner, Nucl. Phys. {\bf B324} (1989)
427.} 
\lref\twisted{E. Witten. Comm. Math. Phys. {\bf 118} (1988) 411\semi
E. Witten, Nucl. Phys. {\bf B340} (1990) 281\semi
T. Eguchi and S.-K. Yang , Mod. Phys. Lett. {\bf A5} (1990) 1693.}
 \lref\GVW{B. Greene, C. Vafa and N. Warner, Nucl. Phys. {\bf B324} (1989)
371.}
\lref\Grisaru{M. Grisaru, A. Van de Ven and D. Zanon, Phys. Lett. {\bf 173B}
(1986) 423.}
\lref\DSWW{M. Dine, N. Seiberg, X. G. Wen and E. Witten, Nucl. Phys. {\bf
B278} (1986) 769, {\bf B289} (1987) 319.}
\lref\DixonRev{L. Dixon, in ``Proceedings of the 1987 ICTP Summer Workshop in
High Energy Physics and Cosmology", ed G. Furlan, {\it et. al.}}
\lref\Exact{J. Distler and B. Greene, Nucl. Phys. {\bf B309} (1988) 295.}
\lref\twozero{J. Distler and B. Greene, Nucl. Phys. {\bf B304} (1988) 1.}
\lref\AspMor{P. Aspinwall, and D. Morrison, ``Topological field theory and
rational curves", Oxford and Duke preprints OUTP-91-32P, DUK-M-91-12 (1991).}
\lref\Morrison{D. Morrison, ``Picard-Fuchs equations and mirror maps for
hypersurfaces", Duke Math preprint DUK-M-91-14 (1991).}
\lref\CandMir{P. Candelas, X. de la Ossa, P. Green and L. Parkes, Nucl.
Phys. {\bf B359} (1991) 21.}
\lref\Kutconf{D. Kutasov, ``Geometry on the space of conformal field
theories and contact terms,'' Phys. Lett. {\bf B220} (1989) 153.}%
\lref\GrSei{M. Green and N. Seiberg, ``Contact interactions in
superstring theory," Nucl. Phys. {\bf B299} (1988) 559.}%
\lref\WilZee{F. Wilczek and A. Zee, Phys. Rev. Lett. {\bf 52} (1984) 2111.}
\lref\Banks{A. Sen, Nucl. Phys. {\bf B278} (1986) 289.\semi
T. Banks, L. Dixon,
D.Friedan and E. Martinec, Nucl. Phys. {\bf B299} (1988) 613.}
\lref\Hodge{P. Griffiths, Periods of Integrals on Algebraic manifolds I,II,
Am. J. Math. {\bf 90} (1970) 568,805\semi
R. Bryant and P. Griffiths, in ``Progress in Mathematics {\bf 36}"
(Birkh\"auser, 1983) 77.}
\lref\Odake{S. Odake, Mod. Phys. Lett. {\bf A4} (1989) 557\semi
S. Odake, Int. Jour. Mod. Phys. {\bf A5} (1990) 897\semi
T. Eguchi, H. Ooguri, A. Taormina and S-K. Yang, Nucl. Phys. {\bf B315}
(1989) 193.}
\lref\GSW{M. Green, J. Schwarz and E. Witten, ``Superstring theory, vol. II"
(Cambridge University Press, 1987).}
\lref\CHSW{P. Candelas, G. Horowitz, A. Strominger and E. Witten, Nucl. Phys.
{\bf B258} (1985) 46.}
\lref\NR{L. Alvarez-Gaum\'e, S. Coleman and P. Ginsparg, Comm. Math. Phys.
{\bf 103} (1986) 423.}
\lref\manin{D. Leites, ``Introduction to the theory of supermanifolds", Russ.
Math. Surveys {\bf 35} (1980) 3\semi
Yu. Manin, ``Gauge field theory and complex geometry", (Springer, 1988).}
\lref\Reviews{
J. Schwarz, ``Superconformal symmetry in string theory", lectures
at the 1988 Banff Summer Institute on Particles and Fields (1988)\semi
D. Gepner, ``Lectures on N=2 string theory",
lectures at the 1989 Trieste Spring School (1989)\semi
N. Warner, ``Lectures on N=2 superconformal theories and singularity theory",
lectures at the 1989 Trieste Spring School (1989)\semi
B. Greene, ``Lectures on string theory in four dimensions",
lectures at the 1990 Trieste Spring School (1990)\semi
S. Yau (editor), ``Essays in Mirror Manifolds",  Proceedings of the Conference
on Mirror Symmetry, MSRI (International Press, 1992).
}
\lref\Seiberg{N. Seiberg, Nucl. Phys. {\bf B303} (1988) 286.}
\lref\Dubrovin{B. Dubrovin, ``Geometry and integrability of
topological--antitopological fusion", INFN preprint, INFN-8-92-DSF (1992).}
\lref\GPM{B. Greene, D. Morrison, and R. Plesser, in preparation.}
\lref\GPmirror{B. Greene and R. Plesser, Nucl. Phys. {\bf B338} (1990) 15.}

\centerline{\bf Notes on N=2 \sm s${}^\star$}\smallskip
{\baselineskip=14pt
\centerline{Jacques Distler}
\centerline{Joseph Henry Laboratories}
\centerline{Princeton University}
\centerline{Princeton, NJ \ 08544 \ USA}
\bigskip\bigskip

\footnote{}{{\parindent=-10pt\par $\star$
\vtop{
\hbox{Research supported by NSF grant PHY90-21984.}
     }     }}
\parindent=20pt

\narrower\narrower\noindent
These lectures are devoted to some
selected topics in N=2 \sm s on Calabi-Yau manifolds and the associated
N=2 superconformal field theories. \myab

}
 \newsec{Introduction}

The subject of N=2 $\sigma$-models and N=2 superconformal field theory is
simply too vast to be adequately treated in these brief lecture notes. A truly
comprehensive review would be a book in itself. Instead of trying to be
comprehensive, I will focus on two topics which I hope will lead the student
to a better appreciation of this rich subject. At the same time, I hope my
slightly unconventional treatment of these topics will be of some interest to
the ``expert". Necessarily, I will leave out much that should be  said.
Luckily, two of the other lectures in this volume, those by Cecotti and
Candelas, will fill in some of the gaps. Also, there are some truly excellent
reviews already existing \refs{\DixonRev,\Reviews,\GSW}.

Supersymmetric \sm s on Calabi-Yau manifolds, aside from any intrinsic
mathematical interest, are usually considered as  backgrounds for
``compactified" string theories. If one wants to get a fermionic string
theory with four noncompact dimensions ($c=6$), one replaces six of the ten
flat dimensions with a $c=9$ superconformal field theory. So, for string
theory, we are really interested in the conformally invariant theory, \ie\ the
 Calabi-Yau \sm\ at its infrared fixed point.

Actually, fixed ``point" is a misnomer. Even at the fixed point, there
generally exist some number of exactly marginal couplings, so the fixed point
set is actually a finite-dimensional variety, ``the moduli space of the
conformal field theory"

 Except, perhaps, at
some exceptional points in the moduli space, we don't really know how to
construct the conformal theory directly. There are lots of different
approaches one can take to overcome this problem. They tend to fall into
two broad classes. One is to try to do what one can, exploiting the rich set
of symmetries imposed by the superconformal invariance to prove certain
things about the CFT. This turns out to be surprisingly powerful. The other
alternative is to study the nonconformal \sm\ which flows to the conformal
theory in the infrared. Clearly, if we can find renormalization
group-invariant quantities, we can study them in the massive theory (where
they are generally easier to calculate), and thereby learn something about the
CFT. The most successful approach is, of course, to combine {it both}
techniques: use conformal methods to uncover symmetries of the theory which
make it {\it easier} to calculate quantities of interest in the noncritical
theory.

One of the most striking examples of this is mirror
symmetry (the subject of Candelas' lectures). There, one finds a nontrivial
automorphism of the conformal field theory which allows one to relate
quantities which have very different \sm\ interpretations. This opens up to
direct \sm\ calculation quantities which previously would have been very
difficult to calculate. I will leave the details for Candelas to unfold, but
I will at least lay the groundwork for some of his calculations.

A word about supersymmetry: the massive theories we are considering have N=2
supersymmetry. At the conformal point, this symmetry is extended to N=2
superconformal symmetry. Actually, one has independent left- and right-moving
superconformal symmetries, so one often says that such theories possess
(2,2) superconformal symmetry. In order to couple one of these theories to
heterotic string theory, it must possess at least (1,0) supersymmetry. In
order to perform a chiral GSO projection, we need an operator $(-1)^{F_L}$
which anticommutes with the left-moving supercurrent, and squares to 1. In a
theory with (2,0) supersymmetry (and integer U(1) charges) there is a natural
candidate for this operator, namely let $F_L= Q=\oint J$, the conserved charge
associated to the $U(1)$ current $J$ in the N=2 superconformal algebra. With
this definition, the theory naturally has {\it spacetime} supersymmetry
\refs{\Banks,\DixonRev} because the states of the Ramond sector are related to
those of the NS sector by spectral flow.

We could study models with just (2,0) supersymmetry. These are {\it very}
interesting from the point of view of the phenomenology of the resulting
compactified string theories. They also pose some outstanding theoretical
challenges. However, unlike the (2,2) models we will study, there is no
natural way to consider them as the IR limits of massive ``(2,0)" theories.
At present, our only tool for studying them is to try to  construct them
directly at the conformal point. This is hard, and as a consequence, little
is known about such theories in general. For the purposes of these notes, I
will stick to the more frequently trodden path of (2,2) supersymmetry.

Section 2 of these notes is devoted to studying the moduli space of these
(2,2) superconformal field theories. In keeping with my description above of
the various ``approaches" to this subject, we will hew to the superconformal
approach, and study directly the properties of the superconformal theory. I
will rely heavily on the calculations of \DKL, who have done the hard
work of calculating the relevant superconformal correlators. Our main focus
will be on the intrinsic geometry of the moduli space, but the main result --
applicable in string theory -- is a determination of the Zamolodchikov
metrics  for
various fields (\ie\ the kinetic terms in the effective spacetime field
theory).

In section 3, I will do an about-face, and take the approach of trying to
find RG-invariant quantities to study in the nonconformal \sm. For
concreteness, we will focus on the cubic couplings which entered into the
formul\ae\ of section 2. These are also of ``phenomenological" interest, for
they determine the superpotential of the effective spacetime theory.

\newsec{N=2 superconformal symmetry and special geometry}

One approach to this subject is to concentrate on the properties of the
theory that follow from the N=2 superconformal symmetry. For an excellent
review of the subject in this spirit, see \DixonRev. Here we will
unravel the geometry of the moduli space of (2,2) superconformal theories
which follow from (2,2) superconformal invariance.

The N=2 superconformal algebra is
\eqn\nalg{\eqalign{T(z)T(w)&={c/2\over(z-w)^4} +{2T(w)\over (z-w)^2}+{\del_w
T\over z-w}\cr
J(z)J(w)&={c/3\over (z-w)^2}\cr
G^+(z)G^-(w)&={c/3\over(z-w)^3}+{J(w)\over(z-w)^2}+{T(w)+\half\del_wJ\over
z-w}\cr
T(z)G^\pm(w)&={3\over2}{G^\pm(w)\over (z-w)^2}+{\del_wG^\pm\over z-w}\cr
J(z)G^\pm(w)&=\pm{G^\pm(w)\over z-w}\cr}}
Primary states of the theory can be classified by their conformal weight
$\Delta$ and their $U(1)$ charge $Q$. Unitarity of the theory requires
\eqn\untara{\Delta\geq \half |Q|}
A primary state $\ket{\phi}$ which violated this inequality would lead to a
negative norm state $G^\pm_{-1/2}\ket{\phi}$. There are a similar set of
inequalities which prevent the occurrence of negative-norm states at higher
levels. Together, these inequalities etch out a convex polygon in the
($\Delta,Q$)-plane whose vertices lie on the envelope
$$\Delta= {3\over2c} Q^2$$

Fields which saturate the bound \untara\ are called chiral primaries \LVW.
They have a null state at the first level $G^\pm_{-1/2}\ket{\phi}=0$.
Putting together left- and right-movers, we have  various possibilities, which
we label as follows:
\item{$\bullet$}chiral primary fields have  $Q=  \Delta/2$ $\bar
Q=\bar\Delta/2$,
\item{$\bullet$}twisted-chiral primaries have $Q=-\Delta/2$
$\bar Q=\bar\Delta/2$.

\noindent Adjoints of these fields (``antichiral" and ``twisted antichiral")
have opposite charge assignments. Unitarity implies the largest allowed charge
for (twisted) chiral field is $|Q_{max}|=c/3$. We demand that there be a
holomorphic field with this charge $\epsilon^\pm(z)$ $Q=\pm c/3$,
$\Delta=c/6$, $\bar Q=\bar\Delta=0$. Clearly, $c$ must be a multiple of 3, so
that the $\epsilon^\pm(z)$ are mutually local. Also, all of the other $U(1)$
charges in the theory must be integral, so that $\epsilon^\pm(z)$ are local
with respect to those fields. When embedded in a string theory, a (2,2)
superconformal theory of this sort leads to a string theory with spacetime
supersymmetry \refs{\DixonRev,\Banks}. The existence of extra holomorphic
fields means our theory has an extended chiral algebra. For $c=6$, this is
simply the  N=4 superconformal algebra, with $\epsilon^\pm, J$ forming the
$SU(2)$ Ka\v c-Moody algebra contained in the N=4 algebra.
 For $c=9$, the extended algebra one gets \Odake\ has no commonly accepted
name.
It contains two non-commuting N=2 algebras, the usual one with supercurrents
$G^\pm(z)$, and a second one with supercurrents $\epsilon^\pm(z)$, U(1)
current $\tilde J={\textstyle{1\over 3}}J$ and stress tensor $\tilde
T={\textstyle{1\over 6}}:J^2:$\ .

We will consider $c=9$ for definiteness. We can organize \refs{\LVW,\Odake}\
the chiral fields of
the model into those which are chiral both with respect to {\it  both} of the
N=2's in the extended chiral algebra, and those which are descendent with
respect to the ``exotic" N=2 generated by $\epsilon^\pm$.

The chiral primary ${\bi h}$ with $Q=\bar Q=1$ has the property that $\ket{h}$
is annihilated by $\epsilon^+_{-1/2}$ and $\bar\epsilon^+_{-1/2}$, so it is a
chiral primary with respect to both N=2's. Its adjoint, ${\bi h}^\dagger$ has
$Q=\bar
Q=-1$ and is antichiral with respect to both N=2's. The twisted chiral primary
${\bi b}$ with $Q=-1$, $\bar Q=1$ has  $\ket{b}$  annihilated by
$\epsilon^-_{-1/2}$
and $\bar\epsilon^+_{-1/2}$ and so is twisted-chiral with respect to both
N=2's. ${\bi b}^\dagger$ has $Q=1$, $\bar Q=-1$.

The rest of the chiral and twisted-chiral primaries (and their adjoints) are
obtained by acting on these with $\epsilon^\pm_{-1/2}$ and
$\bar\epsilon^\pm_{-1/2}$. For instance, we have the chiral primary $\tilde
{\bi h}=(\epsilon^+_{-1/2}\bar\epsilon^+_{-1/2}{\bi h}^\dagger)$ which is the
primary
field associated to the state $\ket{\tilde
h}=\epsilon^+_{-1/2}\bar\epsilon^+_{-1/2} \ket{h^\dagger}$  which has $Q=\bar
Q=2$. In the tables below I have listed the chiral primaries and twisted
chiral primaries and their charges.

\def\tablerule{\omit&&&\multispan{11}{\tabskip=0pt\hrulefill}\cr}
\def\tablepad{\omit&&&height3pt&&&&&&&&&&\cr}
$$\ifx\answ\bigans\else\hskip-.375in\fi\vbox{\offinterlineskip\tabskip=0pt
\halign to 190pt{
    \strut#& #\tabskip=0em plus 1em&\hfil$#$&\vrule#&     \hfil#&
\vrule#& \hfil$#$\hfil&\vrule# &\hfil$#$\hfil& \vrule#&
\hfil$#$\hfil&\vrule# &\hfil$#$\hfil& \vrule#\tabskip=0pt\cr
&&\multispan3&\multispan{9}\hfil $Q \to$\hfil\cr
\omit&&\multispan3&\multispan{9}{\tabskip=0pt\hrulefill}\cr
&&\multispan3&&0&&1&&2&&3&\cr\tablerule\tablepad
&&&&0&&\BI&&&&&&\epsilon^+&\cr\tablerule\tablepad
&&\bar Q&&1&&&&{\bi h}&&(\epsilon_{-1/2}^+{\bi b})&&&\cr\tablerule\tablepad
&&\downarrow&&2&&&&(\bar\epsilon^+_{-1/2}{\bi b}^\dagger)&&
(\epsilon^+_{-1/2}\bar\epsilon^+_{-1/2}{\bi
h}^\dagger)&&&\cr\tablerule\tablepad
&&&&3&&\bar\epsilon^+&&&&&&\epsilon^+\bar\epsilon^+&\cr\tablerule
\noalign{\bigskip}
&\multispan{13}\hfil Chiral Primaries\hfil\cr}}
\hskip.125in\vbox{\offinterlineskip\tabskip=0pt
\halign to 190pt{
    \strut#& #\tabskip=0em plus 1em&\hfil$#$&\vrule#&     \hfil#&
\vrule#& \hfil$#$\hfil&\vrule# &\hfil$#$\hfil& \vrule#&
\hfil$#$\hfil&\vrule# &\hfil$#$\hfil& \vrule#\tabskip=0pt\cr
&&\multispan3&\multispan{9}\hfil $Q \to$\hfil\cr
\omit&&\multispan3&\multispan{9}{\tabskip=0pt\hrulefill}\cr
&&\multispan3&&0&&-1&&-2&&-3&\cr\tablerule\tablepad
&&&&0&&\BI&&&&&&\epsilon^-&\cr\tablerule\tablepad
&&\bar Q&&1&&&&{\bi b}&&(\epsilon_{-1/2}^-{\bi h})&&&\cr\tablerule\tablepad
&&\downarrow&&2&&&&(\bar\epsilon^+_{-1/2}{\bi h}^\dagger)&&
(\epsilon^-_{-1/2}\bar\epsilon^+_{-1/2}{\bi
b}^\dagger)&&&\cr\tablerule\tablepad
&&&&3&&\bar\epsilon^+&&&&&&\epsilon^-\bar\epsilon^+&\cr\tablerule
\noalign{\bigskip}
&\multispan{13}\hfil Twisted Chiral Primaries\hfil\cr}}
$$

  Exactly marginal operators are easy to construct in N=2 theories.
 $C=(G^-_{-1/2}\bar G^-_{-1/2}{\bi h})$, and  $R=(G^+_{-1/2}\bar
G^-_{-1/2}{\bi b})$ are neutral, dimension (1,1) Virasoro primaries\foot{In the
context of N=2 \sm s to be discussed in the next section, these are,
respectively, the moduli corresponding to deformations of the complex structure
and the \Ka\ class of the Calabi-Yau manifold.}.
The primary fields $C,R$ (and their adjoints
$C^\dagger,R^\dagger$) transform into total derivatives under supersymmetry. So
$\int C$ and $\int R$ are superconformally-invariant operators.  We would like
to contemplate adding them to the action
\eqn\Pert{S(\tau,t)=S_0 + \tau^a\int C_a+\bar
\tau^a\int (C^\dagger)_{\bar a}+t^m\int R_m +\bar t^a\int (R^\dagger)_{\bar
m}}
 and thereby defining the deformed theory, whose partition
function $Z(\tau,t)=\int  e^{-S(\tau,t)}$, say, can be calculated
 as a power series expansion in the $\tau$'s
and $t$'s.

Although \Pert\ does indeed define a family of new (2,2) superconformal
theories parametrized by the moduli $\tau,t$, the idea of constructing the
deformed theory as a power series is, unfortunately, a complete fake.

The problem is simple. Each term in the power series expansion is a correlation
function of a product of $C$'s $C^\dagger$'s,
$R$'s and $R^\dagger$'s, each integrated over the world-sheet. But in general,
there are singularities when these operators collide. These singularities
render the integrals ambiguous, and could potentially even ruin the
superconformal invariance of the theory \GrSei.  One prescription for evading
these ambiguities (and showing that (2,2) superconformal invariance is indeed
preserved) is to embed the theory in a string theory. Then the marginal
operators in question are the ``internal" parts of vertex operators for
spacetime fields. We can then consider the correlation function of vertex
operators in a regime in momentum space where the integrals converge, and
{\it define} the full correlation function by analytic continuation in the
momenta. This provides a physical way to understand (and subtract off)
the singularities of the
integrand -- they are the residues of on-shell poles due to intermediate
massless string states. This program was carried out to fourth order by Dixon,
Kaplunovsky and Louis \DKL. The calculation is quite tedious, and probably too
unwieldy to carry out to higher order. Still, as we shall see, it is enough
to recover the local geometry of the moduli space. Indeed, what I am going to
do in the rest of this section is ``borrow" their results. All of the
formul\ae\
which follow appear (perhaps somewhat disguised) in their paper.

Even after eliminating the divergences, one is faced with an ineluctable
fact: the moduli space is a curved manifold. The marginal operators we have
written down represent infinitesimal perturbations \ie\ they are tangent
vectors to the moduli space. To perturb the theory to higher order, we need to
take covariant derivatives, rather than ordinary derivatives. The procedure
used by DKL amounts to a {\it definition} of the connection
one is supposed to use to define these covariant derivatives. Other ways of
subtracting the short-distance singularities of the
correlation-functions lead, in principle, to different definitions of the
connection \Kutconf. Clearly, if our ultimate interest is in embedding these
theories in a string theory, the DKL definition is a good one to follow.

 It is a simple fact of life that on a curved manifold, covariant
derivatives do not commute, and the result of defining the theory as a
function of the moduli by ``parallel transport" is path-dependent\foot{Of
course,  these complications don't arise for a 1-parameter family of
perturbations. One also doesn't notice them at low orders in the perturbation
expansion. Finally, in the particular context in which we are working, {\it
holomorphic} covariant derivatives, $\nabla_{a}$ and $\nabla_{m}$ all
commute. This can fool one into the false impression that the moduli space is
flat.}.

Since the moduli space is curved, when we parallel transport the theory around
a closed loop, we return to an isomorphic, but not quite identical conformal
field theory. In general, the states at each level return to themselves
permuted by a $U(n)$ transformation. This phenomenon, in the much simpler
context of nonrelativistic quantum mechanics is known as {\it Berry's Phase}
(or its nonabelian generalization \WilZee). In particular, the chiral
primaries ${\bi b}$ and ${\bi h}$ come back mapped nontrivially among
themselves. Even the
generators of the $c=9$ extended chiral algebra are  affected, but
there the effect is rather tightly constrained, since they must satisfy the
same algebra as before. The result is that the N=2 supercurrents $G^\pm$
and $\epsilon^\pm$ (and the corresponding barred fields) come back to
themselves multiplied by phases
\eqn\AlgBerry{\eqalign{
G^\pm(z)\to e^{\pm i\theta}G^\pm(z)&,\qquad \bar
G^\pm(\bar z)\to e^{\pm i\tilde \theta} G^\pm(\bar z)\cr
\epsilon^\pm(z)\to e^{\mp3i\theta}\epsilon^\pm(z)&,\qquad\bar\epsilon^\pm(\bar
z)\to
e^{\mp3i\tilde\theta}\bar \epsilon^\pm(\bar z)\cr}}
The stress tensor $T$ and $U(1)$ current $J$ (and their barred
counterparts) must come back to themselves, with no phases because of the
central terms in the algebra \nalg.

When discussing a family of theories, then, we should think of $G^+$ not as a
fixed operator, but as an operator-valued section of a line bundle $L$.
The Berry phase that we pick up is simply a manifestation of the curvature of
$L$.  $G^-$ is an operator-valued section of the dual line bundle
$L^{-1}$. Similarly,  $\bar G^\pm$ are
 operator-valued sections of line bundle $\tilde L,\tilde L^{-1}$ and
$\epsilon^\pm$ are sections of $L^{-3},L^3$ (the third tensor powers of
$L^{-1},L$), \etc

We have already identified the marginal operators
$C$ and $R$ as operator-valued sections of $T$, the (holomorphic) tangent
bundle to the moduli space. Therefore we learn that the chiral primaries
${\bi h}$ and ${\bi b}$ are, respectively, sections of $T\otimes L\otimes\tilde
L$ and
$T\otimes L^{-1}\otimes\tilde L$. In particular, the
precise mapping between the chiral primaries ${\bi b},{\bi h}$ and the moduli
$C,R$ for a
family of theories depends on a choice of section $s\in\Gamma(L)$ and
$\tilde s\in\Gamma(\tilde L)$. To jump slightly ahead, this means that the
Zamolodchikov metric for the chiral primaries {\it differs} from that of the
moduli ({\it c.f.}~\DKL, eqn. 3.36)
\eqn\metrel{\CG_{a\bar
b}=\vev{{\bi h}_a(1)({\bi h}^\dagger)_{\bar b}(0)}=\|s\|^2\|\tilde s\|^2
g_{a\bar b}}
where
$$g_{a\bar b}=\vev{C_a(1)(C^\dagger)_{\bar b}(0)}$$
(with a similar formula for ${\bi b}$'s and $R$'s).

Let me repeat that we're not
saying anything exotic here. We started out with states $\ket{C_a}$ and
$\ket{h_a}$ related by $\ket{C_a}=G^+_{-1/2}\bar G^+_{-1/2}\ket{h_a}$. We
parallel transport them around a closed loop in moduli space, and they
come back rotated by a $U(n)$ matrices, $\ket{C_a}\to U_a{}^b\ket{C_b}$,
$\ket{h_a}\to U'_a{}^b\ket{h_b}$. Superconformal symmetry dictates that the
two $U(n)$ matrices must be related:
$U_a{}^b=e^{i(\theta+\tilde\theta)}U'_a{}^b$ and that, moreover, the Berry
phases for {\it all} of the states in the superconformal module built on $
\ket{h_a}$ must be similarly related.

To understand the Berry's phase in this sector of the theory, we need to
find the metric on moduli space, and
the fiber metric on the line bundle $L$ and $\tilde L$.

 The trick to constructing these objects is to  realize that
the same moduli space
 parametrizes the {\it topological}
field theories obtained by twisting this N=2 superconformal theory\twisted. The
topological theory is obtained by ``improving" the stress tensor $T\to
T\pm\half \del_zJ$ (and similarly for $\bar T$) so that one of the
supercurrents
$G^\pm$ becomes dimension 1, while the other becomes dimension 2. The integral
of the dimension 1 supercurrent then becomes a nilpotent global charge $Q$,
and we {\it define} the physical states of the topological theory to be those
in the cohomology of $Q$. There are two independent twistings of relevance:
the ``A" model, in which  $Q=\oint G^-$ and $\bar Q=\oint \bar G ^+$ are the
nilpotent charges which define the theory, and the ``B" model, in which
$Q=\oint G^+$ and $\bar Q=\oint \bar G ^+$ are the nilpotent supercharges
\AandB. In each case, the physical states of the topological theory are in 1-1
correspondence with the twisted chiral or chiral primaries listed
above.

In the topological theory, charge is violated by 3 units on the sphere. So
the two-point function provides a quadratic form on the (twisted) chiral ring
$$\eta(\phi_i,\phi_j)=\vev{\phi_i(z,\bar z)\phi_j(0)}$$
where, in order to get a nonzero 2-point function, the total charge must add up
to $\bar Q=3$ and $Q=\pm3$. This quadratic form is symmetric or
skew-symmetric\foot{The skew-symmetric part of this quadratic form for the
twisted chiral ring (\ie\ the quadratic form restricted to the
skew-diagonal part of the above table of twisted chiral primaries) is the
conformal field theory version of the quadratic form studied in the
``variation of Hodge structures" approach to the moduli space of Calabi-Yau
manifolds \refs{\Hodge,\StromSpec}.}, depending on whether the total charge
($Q+\bar
Q$) of $\phi_i$
 is even or odd. (By topological invariance, the 2-point functions is
actually independent of $z,\bar z$.)

Demanding that the connection be compatible both with the metric (of the
tangent bundle and of the line bundles $L,\tilde L$) and with this quadratic
form is extremely restrictive\foot{For a discussion of the
topological-antitopological fusion equations as compatibility equations, see
\Dubrovin.}. It implies a set of equations which allows us
to solve for the metrics. How exactly one phrases these equations is somewhat
a matter of taste. They appear in the theory of variations of Hodge
structures \Hodge, and in the Calabi-Yau context are quite elegantly discussed
in \StromSpec. Slightly repackaged in the context of massive deformations of
N=2 theories, they go by the name of ``topological-antitopological fusion"
equations \TopAntiTop, which are discussed by Cecotti in his lectures.
Alternatively, one can simply calculate the relevant 4-point functions in the
superconformal field theory \DKL.

However one arrives at it, the result is simple to state. The line bundles
$L,\tilde L$ have curvature tensors $F,\tilde F$ which are both of type (1,1).
What is more, the mixed components of these curvature tensors vanish
\eqn\Fspeca{F_{a\bar m}=F_{m\bar a}=\tilde F_{a\bar m}=\tilde F_{m\bar a}=0}
and the remaining components satisfy
\eqn\Fspecb{F_{a\bar b}=\tilde F_{a\bar b},\qquad F_{m\bar n}=-\tilde F_{m\bar
n}}
The line bundles $L,\tilde L$ can actually be given the structure of
holomorphic line bundles, and the DKL definition corresponds to choosing a
holomorphic connection on $L,\tilde L$.

 The metric on the moduli space also takes a block-diagonal form:
\eqn\gspec{g_{a\bar b}=3 \tilde F_{a\bar b},\qquad g_{m\bar n}=3\tilde
F_{m\bar n}} 
This means the moduli space is \Ka\ since $\tilde F$ is a closed 2-form, and
\Fspeca,\Fspecb\ imply that the mixed components of Christoffel connection,
like $\Gamma ^a_{mb}$, and the mixed components of the curvature vanish. This
is not quite enough to prove that the moduli space is a product manifold,
${cal M}\times{cal M}'$, parametrized by the $\tau$'s and $t$'s respectively. In many
cases of interest, it appears that the moduli space is the quotient of such a
product manifold by the action of a discrete group which acts nontrivially on
both factors. We will ignore this subtlety, assuming that we can always go to
a suitable covering space which {\it is} a product ${cal M}\times{cal M}'$.

There is one more condition which restricts the geometry of the moduli space,
but to state it, we must pause to  introduce the 3-point functions (the
``(twisted) chiral ring" \LVW, or the ``Yukawa couplings" of the low-energy
effective field theory \GSW)
$$\eqalign{W_{abc}(s^{\otimes 6}\otimes \tilde
s^{\otimes 6})&=\vev{\tilde {\bi h}_a(0)
{\bi h}_b(1){\bi h}_c(-1)} \cr W'_{kmn}((s^{-1})^{\otimes 6}\otimes \tilde
s^{\otimes 6})&=\vev{\tilde
{\bi b}_k(0) {\bi b}_m(1){\bi b}_n(-1)} \cr}$$
where $\tilde {\bi h}_a=(\epsilon_{-1/2}^-\bar \epsilon_{-1/2}^- {\bi h}_a)$,
and
$\tilde {\bi b}_m=(\epsilon_{+1/2}^-\bar \epsilon_{-1/2}^- {\bi b}_m)$. The
3-point
functions depend explicitly on the sections $s,\tilde s$ which relate the
tangent vectors to the moduli space to the chiral primaries ${\bi h},{\bi b}$
(with some
extra powers of $s,\tilde s $ thrown in to define the $\epsilon$'s. {\it
c.f.}~\AlgBerry).

$W_{abc}$ and $W'_{kmn}$ don't look symmetric in their
indices, but by the Ward identities associated to the extended $c=9$ chiral
algebra, they actually are symmetric\foot{The Ward
identity in question simply says that $$\vev{\oint \dd z (z-w)/w \
\epsilon^-(z) \hat {\bi h}_a(0) {\bi h}_b(1) {\bi h}_c(-1)},$$ where the
contour surrounds the
origin, is independent of $w$. Doing the same contour-deformation trick
discussed below for both $\epsilon^-$ and $\bar\epsilon^-$, we conclude
that $W_{abc}$ is independent of which operator has the ``tilde" on
it.}. So $W$ is a section of $S^3(T^*)\otimes L^{-6}\otimes\tilde L^{-6}$ and
$W'$ is a section of $S^3(T^*)\otimes L^{6}\otimes\tilde L^{-6}$. They are
actually holomorphic sections:
\eqna\Wspec$$\eqalignno{\nabla_{\bar m} W_{abc}&=\nabla_{\bar
d}W_{abc}=0&\Wspec a\cr
\nabla_{\bar m}W'_{kmn}&=\nabla_{\bar a}W'_{kmn}=0&\Wspec b\cr}$$
The proof of this is an easy application of the superconformal Ward
identities \Exact. Because of the \Ka\ geometry, the Christoffel terms vanish
and
$$\nabla_{\bar m}W_{abc}=\del_{\bar m}W_{abc}=\vev{\tilde
{\bi h}_a(0){\bi h}_b(1){\bi h}_c(-1)\int\dd^2wR^\dagger_{\bar m}(w,\bar w)}$$
Write
$$\int\dd^2w R^\dagger(w,\bar w) =-\int\dd^2w \oint\dd \bar y {\bar
y-\bar u\over \bar w-\bar u}\bar  G^+(\bar y) \ (G^-_{-1/2}{\bi
b}^\dagger)(w,\bar
w)$$ where the contour surrounds $w$. The Ward identity says the
correlation function is actually independent of $\bar u$. Setting $\bar u=0$,
one can deform the contour so that it closes around the other vertex
operators. One gets no contribution from closing the contour around $\pm1$,
since $\bar G^+_{-1/2}{\bi h}=0$. The contribution from $0$ vanishes as well,
since
$\bar G^+$ has only a single pole with the operator at the origin, and there is
an explicit factor of $\bar y$ in the integrand.

For $\vev{\tilde {\bi h}_a(0){\bi h}_b(1){\bi h}_c(-1)\int
\dd^2wC^\dagger_{\bar d}(w,\bar w))}$,
the argument goes almost the same way, except that we have to worry a little
more about short-distance singularities. However, because we can deform both
the $G^+$ and the $\bar G^+$ contours, we can show that the residues
actually vanish.

We would like to use identical reasoning to conclude that the mixed
holomorphic covariant derivatives also vanish
$$\eqalignno{\nabla_mW_{abc}&=\nabla_aW'_{kmn}=0&\Wspec c\cr}$$
It is certainly true that $(L\otimes\tilde L)|_{{cal M}'}$ is flat, but it is not
necessarily holomorphically trivial. If it is not, we can find locally
covariantly constant sections which have global monodromies. This will mean
that $W$
will also have global monodromy: As we go around a closed loop in ${cal M}'$, $W$
will pick up a phase. This phase  cancels in the expression below for the
curvature of the moduli space. Modulo this subtlety, we simply need to
establish
$$\vev{\tilde {\bi h}_a(0) {\bi h}_b(1){\bi h}_c(-1)\int \dd^2w R_m(w,\bar
w)}=0$$
which again follows from the superconformal Ward identities \Exact.

The final condition that the curvature tensor of the metric $g$ must satisfy is
\eqn\Rspec{\eqalign{R_{a\bar cb\bar d}&=g_{a\bar c}g_{b\bar
d}+g_{a\bar d}g_{b\bar c}
-\|s\|^{-12}\|\tilde s\|^{-12}W(s^6\otimes\tilde s^6)_{abe}
\overline{W(s^6\otimes\tilde s^6)}_{\bar f\bar c \bar d}g^{e\bar
f}\cr R_{k\bar m l\bar n}&=g_{k\bar m}g_{l\bar n}+g_{k\bar n}g_{l\bar m}-
\|s\|^{+12}\|\tilde
s\|^{-12}W'(s^{-6}\otimes\tilde s^6)_{klo}
\overline{W'(s^{-6}\otimes\tilde s^6)}_{\bar p\bar m \bar n}g^{o\bar p}\cr}}
(The other components of $R$ must vanish by the above conditions.)

Note that in defining the 3-point functions, we had to make an arbitrary choice
of sections $s,\tilde s$. Under a change of section $s\to f s$, $\tilde s\to
\tilde f\tilde s$, the 3-point functions  transform
$$W_{abc}\to f^6\tilde f^6 W_{abc},\qquad W'_{kmn}\to f^{-6}\tilde
f^6W'_{kmn}$$
But the factors of $\|s\|^{-12}\|\tilde s\|^{-12}$ in \Rspec\ transform in a
compensating way.
Thus  \Rspec\ is a completely covariant equation for the curvature
of the moduli space.  This collection of restrictions \Fspeca--\Rspec{}\ are
collectively known as ``special geometry"
\refs{\SpecialGeo,\ranspec,\CandMod,\DKL,\StromSpec,\PerStrom}.

These formul\ae\ are usually written, locally, in terms of the \Ka\ potential
for the metric $g$, rather than the norm-squared of a section of a line
bundle. The correspondence is easy to establish if one notes that, given a
meromorphic section $\tilde s$ of $\tilde L$, then the curvature $\tilde F$ is
given by
$$\tilde F=-\del\delb \log\|\tilde s\|^2$$
So the \Ka\ potential for the metric $g$ is just
$$K=-3\log\|\tilde s\|^2$$

Several comments are in order. First, a crucial role was played the fact
that the superconformal generators pick up a Berry phase. This is to be
expected on general grounds, but is frequently ignored in discussions of this
subject.

Second, the framework we have developed here is tailor-made to address
questions
about the global properties of the moduli space. Other treatments of special
geometry, written in local ``special coordinates" are incapable of even
detecting these subtleties. To cite two of these subtleties that we have seen
in our discussion, the first was that the moduli space is in general not a
product manifold. We ``solved" this
problem by going to a suitable covering space which {\it is} a product space
${\cal M}\times{cal M}'$. The second subtlety is that the ``chiral ring" W has
monodromies as one goes around loops in ${cal M}'$, whereas most people (myself
included)
have simply assumed that it is independent of the moduli on ${cal M}'$ (and vice
versa for the twisted chiral ring $W'$ and ${cal M}$). This is because the line
bundles $(L\otimes\tilde L)|_{{cal M}'}$ and $(L^{-1}\otimes\tilde L)|_{{cal M}}$ are
flat, but not necessarily holomorphically trivial.

Third, it is clear that, even ignoring the subtle global questions, it is a
bad idea to focus on one set of the moduli and ignore the other in discussing
the physics of these theories. The Zamolodchikov metrics for various descendent
fields (in particular, the fields $(\bar G^-_{-1/2}{\bi b}),\ (\bar
G^-_{-1/2}{\bi h})$
which go into defining the vertex operators for the matter fields in the string
theory) depend on both sets of moduli \DKL. The Zamolodchikov metrics of the
matter fields
 $$\eqalign{G_{a\bar b}&=\vev{(\bar G^-_{-1/2}{\bi h}_a)(1)(\bar
G^+_{-1/2}{\bi h}^\dagger_{\bar b})(0)}=\|s\|^2g_{a\bar b}\cr
G_{m\bar n}&=\vev{(\bar G^-_{-1/2}{\bi b}_m)(1)(\bar
G^+_{-1/2}{\bi b}^\dagger_{\bar n})(0)}=\|s\|^{-2}g_{m\bar n}\cr}$$
depend nontrivially on both sets of moduli
 so the ``physics" of these theories does not factorize, even {\it
locally}.

Fourth, our analysis has shown that the \Ka\ form of the metric
on the moduli space, ${\cal J}={i\over2\pi} (g_{a\bar
b}\dd\tau^a\wedge\dd\bar\tau^b+g_{m\bar n}\dd t^m\wedge\dd \bar
t^n)=3c_1(\tilde L)$ is {\it three} times an integer class. This
result, anticipated in a footnote in \StromSpec, and implicit in the results of
\DKL\ is  a somewhat more restrictive condition than that obtained in most
previous analyses.

Fifth, and
perhaps most important, our ability to determine the geometry of the moduli
space has been reduced to the necessity of calculating the 3-point functions
$W_{abc},W'_{kmn}$, and solving some differential equations. In special
coordinates, the latter reduce to relatively tractable equations.
The key feature which will make it
possible to carry out this program is the equation \Wspec{}, which tell us
about the covariant constancy of the $W$'s with respect to certain of the
moduli. This is a tremendous advantage in calculating them for certain classes
of theories.

\newsec{N=2 sigma model in super-space}

The superconformal approach followed in the last section is very general --
it applies to essentially any family of $c=9$ N=2 conformal field theories with
the extended chiral algebra. In this section we will turn to a particular
class of such theories which arise from N=2 supersymmetric \sm s on Calabi-Yau
manifolds \refs{\CHSW,\GSW}. We'll concentrate on the problem of calculating
the
3-point functions which were the crux of the special geometry discussed in the
last section.

The \sm\ action can be written in N=2 superspace as
\eqn\Ssigma{S=
\int \dd^2z \dd^4\theta K(\Phi,\bar\Phi)}
where
$\Phi^i$  are superfields whose lowest component $\phi^i$ are local complex
coordinates on a compact Calabi-Yau manifold $M$. The $\Phi^i$ obey a chiral
constraint
\eqn\chiconst{D_+\Phi^i=\bar D_+\Phi^i=D_-\bar\Phi^{\ib}=\bar
D_-\bar\Phi^{\ib}=0}
 where the superderivatives are
$$\eqalign{D_+&=\del_{\theta^+}+\theta^-\del_z,\qquad
D_-=\del_{\theta^-}+\theta^+\del_z\cr
\bar D_+&=\del_{\bar \theta^+}+\bar \theta^-\del_{\bar z},\qquad
\bar D_-=\del_{\bar \theta^-}+\bar\theta^+\del_{\bar z}\cr}$$
In components,
$$\Phi^i=\phi^i(y,\yb)+\theta^-\lambda^i(y,\yb)+\bar \theta^-\psi^i(y,\yb)
+\theta^-\bar\theta^-F^i(y,\yb)$$
where $y=z+\theta^-\theta^+$. The lowest component is, as already mentioned,
a scalar which is a local complex coordinate on $M$. $\lambda^i,\psi^i$ are
left- and right-chirality spinors taking values in $T$, the holomorphic tangent
bundle of $M$, and $F^i$ is an auxiliary field which allows the supersymmetry
algebra to close off-shell. The function $K(\phi,\bar\phi)$ is called the \Ka\
potential, and determines the metric on $M$ by
\eqn\Gdef{g_{i\jb}=\del_i\del_{\jb}K}  The \Ka\ potential is not a
globally-defined function on $M$. Across coordinate patches, $K$ transforms by
a K\"ahler transformation $$K'=K+f(\Phi)+\bar f(\bar\Phi)$$ The \Ka\ metric is
unaffected by \Ka\ transformations, and so {\it is} globally well-defined. The
action, too, is globally defined because we can write \Ssigma\ in terms of
the \Ka\ metric as
\eqn\Strue{S=-
 \int \dd^2z \dd\theta^+\dd\bar\theta^-
D_-\Phi^i\bar D_+\bar\Phi^{\jb} g_{i\jb}(\Phi,\bar\Phi)
-
\int \dd^2z \dd\theta^-\dd\bar\theta^+ D_+\bar\Phi^{\jb}\bar
D_-\Phi^i g_{i\jb}(\Phi,\bar\Phi)}

Clearly, the action can be generalized by adding a $\theta$-term. Let $B$ be
a real closed 2-form on $M$. Add to the action
\eqn\Stheta{\eqalign{S_\theta&=i\int\phi^*B=i\int B_{i\jb}\dd\phi^i\wedge
\dd\phi^{\jb}\cr
&=-\int\dd^2z\dd\theta^+\dd\bar\theta^-
D_-\Phi^i\bar D_+\bar\Phi^{\jb} B_{i\jb}(\Phi,\bar\Phi)
+
\int \dd^2z \dd\theta^-\dd\bar\theta^+ D_+\bar\Phi^{\jb}\bar
D_-\Phi^i B_{i\jb}(\Phi,\bar\Phi)}} This is (almost) a total derivative. Its
integral  is only nonzero for topologically nontrivial maps $\phi$. It thus
has no effect in \sm\ perturbation theory, though it will be very important
when we come to discuss \sm\ instantons.

Generically, the \sm\ \Ssigma+\Stheta\ is not conformally invariant. The
nonrenormalization theorem \NR\ is, however a powerful restriction on the
form of the $\beta$-function. At one-loop, one has a counterterm of the form
\eqn\oneloop{\int \dd^2z \dd\theta^+\dd\bar\theta^- D_-\Phi^i\bar
D_+\bar\Phi^{\jb} R_{i\jb}(\Phi,\bar\Phi)
+\int \dd^2z \dd\theta^-\dd\bar\theta^+ D_+\bar\Phi^{\jb}\bar
D_-\Phi^i R_{i\jb}(\Phi,\bar\Phi)}
where $R_{i\jb}$ is the Ricci tensor on $M$. If one uses the
Ricci-flat \Ka\ metric which is known to exist on a Calabi-Yau manifold, this
actually vanishes, but for any $g_{i\jb}$, the cohomology class of the Ricci
form
$$\CR=i R_{i\jb} \dd\phi^i\wedge\dd\phi^{\jb}$$
vanishes, which means that one can rewrite \oneloop\ as the
$\int\dd^4\theta$ of a globally-defined function on $M$ (a ``D term"). The
nonrenormalization theorem says that all higher-loop counterterms are D-terms
\foot{A historical note: it was once believed
that if one could make the one-loop $\beta$-function vanish in this theory,
the $\beta$-function would then be zero to all orders. Later, a 4-loop
contribution to the $\beta$-function was found \Grisaru.}.

Clearly, the D-terms are renormalized in some horribly complicated way along
renormalization group flows.  It
is believed that they are marginally irrelevant, and flow to zero in
the IR. However, those terms in the action which cannot be
written (globally) as D-terms are protected from renormalization, and are
constant along RG flows. On such term has already been alluded to: the
cohomology class of the \Ka\ form
$$J=ig_{i\jb}\dd\phi^i\wedge\dd\phi^{\jb}$$
which enters into \Strue. Shifting the \Ka\ form by an exact form changes
\Strue\ by a D-term.

Another obvious RG invariant is the cohomology class of the 2-form
$B=B_{i\jb}\dd\phi^i\wedge\dd\bar\phi^j$. It is conventional to combine the
two, and call $\omega=[J+iB]\in\H{2}{M,\BC}$ the cohomology class of the
(generalized) \Ka\ form.

These RG invariants characterize the IR fixed point(s) of this \sm.
Thus we have recovered some of the moduli discussed in the last
section.
\eqn\Rdef{\eqalign{R&=\int\dd\theta^+\dd\bar\theta^-D_-\Phi^i\bar
D_+\bar\Phi^{\jb}b_{i\jb}(\Phi,\bar\Phi)\cr
R^\dagger&=\int\dd\theta^-\dd\bar\theta^+D_+\bar\Phi^{\jb}\bar
D_-\Phi^ib_{i\jb}(\Phi,\bar\Phi)\cr}} where $b_{i\jb}$ is a real
closed 2-form on
$M$. Clearly, $R$ shifts both $g_{i\jb}$ and $B_{i\jb}$ by  $b_{i\jb}$,
whereas $R^\dagger$ shifts them in opposite direction. Since these are
composite operators, they are subject to renormalization. However the
cohomology class of $b_{i\jb}$ is an RG invariant.

Are there any other RG invariants of these \sm s? One which has been lurking in
our formalism is the complex structure of $M$. That is what distinguishes the
coordinates $\phi^i$ from $\phi^{\ib}$ which satisfy {\it different}
chirality constraints \chiconst. Since the complex structure is built into
the  N=2 superspace formalism, it is a little awkward to do explicit
finite variations of it. Nevertheless, an infinitesimal variation of the
complex structure is given by $h_{\ib}{}^j$, an element of the cohomology
group $\H{1}{M,T}$. We {\it can} write down the corresponding marginal operator
\eqn\Cdef{C=\int\dd\theta^-\dd\bar\theta^- D_+\bar\Phi^{\bar k}\bar
D_+\bar\Phi^{\jb}g_{j\bar k}(\Phi,\bar\Phi)h_{\ib}{}^j(\Phi,\bar\Phi)}
Again, the form of this operator gets renormalized in some complicated
way along RG flows, but the cohomology class of $h_{\ib}{}^j$ is an RG
invariant, and characterizes the superconformal fixed points.

We have now identified the moduli of the (2,2) superconformal field theory
which is the IR fixed point of this \sm\ as deformations of either the \Ka\
class, or the complex structure of $M$. These are the only deformations which
we
can probe away from the conformal point. There {\it may} be  other marginal
deformations which break the (2,2) supersymmetry of the conformal point down
to something smaller (like, say (2,0)), but, again, that is beyond the scope
of these notes.

Let us now turn to the 3-point functions $W,W'$ which we introduced in the
last section. We have already written down the operators ${\bi h},{\bi b}$ as
(lowest
components of) superfields.
\eqn\hbdef{{\bi h}=D_+\bar\Phi^{\bar k}\bar
D_+\bar\Phi^{\jb}g_{j\bar
k}(\Phi,\bar\Phi)h_{\ib}{}^j(\Phi,\bar\Phi),\qquad
{\bi b}=D_-\Phi^i\bar
D_+\bar\Phi^{\jb}b_{i\jb}(\Phi,\bar\Phi)}
The detailed form of these operators may, as I have said, get renormalized,
but they are constrained to remain chiral and twisted-chiral superfields,
respectively, \ie\  ${\bi h}$ satisfies $D_+{\bi h}=\bar D_+ {\bi h}=0$ and
${\bi b}$ satisfies
$D_-{\bi b}=\bar D_+{\bi b}=0$

We also need the operators $\tilde {\bi h},\tilde {\bi b}$, which which satisfy
the
appropriate antichiral and twisted antichiral constraints. They take the
form
\eqn\twbh{\eqalign{\tilde {\bi h}&=D_-\Phi^j D_-\Phi^k\bar D_-\Phi^{j'}\bar
D_-\Phi^{k'}
\epsilon_{ijk}\epsilon_{i'j'k'} g^{i'\ib}h_{\ib}{}^i\cr
 \tilde
{\bi b}&=D_+\bar\Phi^{\jb}D_+\bar\Phi^{\bar k} \bar D_-\Phi^{j}\bar
D_-\Phi^{k}\epsilon_{\ib \jb\bar k}\epsilon_{ijk}g^{i\bar l}g^{l\ib}b_{l\bar
l}}}
 where $\epsilon_{ijk}$ is the
holomorphic 3-form on $M$ \CHSW.

The 3-point function $W_{abc}$ is then given to lowest order by the supergraph
\tfig\grapha, and $W'_{kmn}$ by \tfig\graphb.

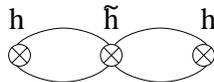
\ifigures{\grapha}{Supergraph
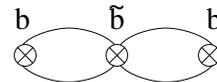
contributing to $W_{abc}$}{grapha.eps}{.492}{\graphb}{Supergraph
contributing to $W'_{kmn}$}{graphb.eps}{.492}

I know of no direct proof that there are no higher-loop corrections to these
graphs (beyond those which renormalize the operators themselves). But the
general arguments lead us to that conclusion. The reason is simple. We
proved that $\nabla_{\bar m}W_{abc}=\nabla_{\bar m}W'_{kmn}=0$. This means
that the 3-point functions depend holomorphically on the generalized \Ka\
class, \ie\ they depend only on the combination $\omega=[J+iB]$. But
perturbation theory is completely insensitive to $B$ because $S_\theta$ is
locally a total derivative. Thus the lowest-order perturbative result must be
exact.

Of course \sm\ instanton corrections {\it do} depend on $B$, so $W'_{kmn}$
does receive instanton corrections \DSWW. However, we have also shown that
$\nabla_mW_{abc}=0$, so $W_{abc}$ must be completely independent of the
generalized \Ka\ class. Hence it receives no instanton corrections and is
entirely given by the lowest order graph \grapha. The result is well-known
\GSW,
$$W_{abc}=\int_M \left(h_{\ib}^{(a)\ i'}h_{\jb}^{(b)\ j'}h_{\bar k'}^{(c)\
k}\epsilon_{i'j'k'}\ \epsilon_{ijk}\right)$$
where the 6-form in parentheses is integrated over the Calabi-Yau manifold
$M$. (One always has this remaining integral to do in background field
perturbation theory.)

The perturbative result for $W'_{mnp}$ is similar\foot{Note that in writing
this result, we are making an implicit choice for the holomorphic 3-form
$\epsilon$ which appeared in \twbh, but has disappeared from this formula.
The choice
which we are making is to take $\epsilon$ to be a generator of integral
cohomology. This makes sense so long as we are holding fixed the complex
structure of $M$. The fact that the formula for $W'_{mnp}$ {\it depends} on
such
a choice was transparent in the formalism of section (2), and has been
independently emphasized in \GPM.}
\eqn\Wclass{W'^{(cl)}_{mnp}
=\int_M\left(b^{(m)}_{i\ib}b^{(n)}_{j\jb}b^{(p)}_{k\kb}\right)
\equiv\int_Mb^{(m)}\wedge b^{(n)}\wedge b^{(p)}}
This, however, {\it does} receive instanton corrections.

The point is that there are nontrivial, finite-action solutions to the
equations
of motion given by solutions to
\eqn\class{\del_{\bar\theta^+}\Phi^i=D_-\bar D_-\Phi^i+
\Gamma^i_{jk}(\Phi,\bar\Phi)D_-\Phi^j\bar D_-\Phi^k=0}
corresponding to holomorphic curves in $M$.

In components, these equations (and their barred counterparts) are
\eqna\comps
$$\eqalignno{
\del_{\zb}\phi^i=\del_z\phi^{\ib}=\del_{\zb}\lambda^i&=\del_z
\psi^{\ib}=0&\comps a\cr
F^i=\Gamma^i_{jk}\lambda^j\psi^k,\qquad F^{\ib}&=-\Gamma^\ib_{\jb\bar
k}\lambda^\jb\psi^{\bar k}&\comps b\cr
\del_z\psi^i+\Gamma^i_{jk}\del_z\phi^j\psi^k&=R^i{}_{jk\bar l}\lambda^{\bar
l}\lambda^j\psi^k&\comps c\cr
\del_{\zb}\lambda^\ib+\Gamma^\ib_{\jb\bar k}\del_\zb\phi^\jb\lambda^{\bar k}&=
R^\ib{}_{\jb\bar k l}\lambda^{\bar k}\psi^l\psi^\jb&\comps d\cr
}$$

Actually, \comps{c,d}, though they exactly stationarize the action, are awkward
to solve. Instead one can solve the linearized equations (\comps{c,d}\ with the
R.H.S.~set equal to zero), and obtain an approximate stationary point of the
action, supplemented by an explicit 4-fermi term. One can further simplify
\comps{c,d}\ by performing a chiral change of variables\foot{nonanomalous
because $c_1(M)=0$.}
$$\lambda^\ib=g^{\ib j}\lambda_j,\qquad\psi^j=g^{\ib j}\psi_\ib$$
The equations \comps{c,d}\ then become
$$\eqalignno{\del_\zb\lambda_j=\del_z\psi_\ib&=0&\comps {c'}\cr}$$
The 4-fermi term in the action is
$$S_4=\int R^{i\jb}{}_{k\bar l}\lambda_i\lambda^k\psi_\jb\psi^{\bar l}$$
Depending on the number of fermion zero modes in the instanton background,
we may have to bring down powers of this term to soak up any ``extra" zero
modes
not absorbed by the operator insertions in the correlation function \DSWW.

To solve \comps{a}, let $C\subset M$ be a holomorphic $\cp{1}$ in $M$. We
obtain a solution if $\phi: \Sigma\to C$ is a holomorphic
map from the world-sheet $\Sigma$ (also a $\cp{1}$) to $C$. In appropriate
local coordinates on $M$ (so that, say, $C$ is described locally as the
$\phi^3$-plane), such a map is just a rational function
\eqn\ratfun{\phi^3(z)={\sum_{i=0}^k a_i z^i\over\sum_{i=0}^k b_i z^i}}
This map has winding number $k$. Clearly, it is
invariant under  a common rescaling  $\{a_i,b_i\}\to \{\lambda a_i,\lambda
b_i\}$. To obtain a smooth map of winding number $k$,  we
should, strictly speaking, demand that the roots of the two polynomials in the
numerator and denominator of \ratfun\ not coincide. However for computing
integrals, it is natural to compactify the space of solutions by relaxing this
restriction, allowing all $\{a_i,b_i\}$ modulo a common rescaling. The
compactified ``instanton moduli space" is thus isomorphic to $\cp{2k+1}$
\AspMor.

To continue with the computation, we need to discuss the fermionic zero modes
in the background \ratfun. Solving \comps{a,{c'}}\ is trivial. The $\lambda$
zero modes are holomorphic, and the $\psi$'s antiholomorphic. The only
restriction
comes from demanding that the zero modes be normalizable. That is to say, the
$\lambda$'s are sections of certain holomorphic line bundles on $\Sigma$, and
these line bundles have only a finite number of global holomorphic sections.
For a holomorphic instanton of winding number $k$ \ratfun, there are\foot{For
the  cognoscenti, I
am assuming that $C$ is an isolated curve of type $(-1,-1)$. That is, I
assume the tangent bundle of $M$, when restricted to $C$ splits as
$TM|_C\simeq\CO(2)\oplus\CO(-1)\oplus\CO(-1)$.}\ $2k$ zero modes of
$\lambda^3$, and $k$ zero modes each of $\lambda_1,\lambda_2$ (with the
corresponding number of zero modes of the $\psi$'s).

For $k=1$, this means
that there are a total of 4 zero modes of $\lambda$, and 4 zero modes of
$\psi$, precisely the number which can be absorbed by the operators in the
expression for $W'_{mnp}$.
Thus we can ignore the 4-fermi term in the action, and the
computation is completely straightforward. The instanton action is just
given by $S_{inst}=\int_C\omega$ The operators absorb the fermion zero modes,
and  the integral over the instanton moduli space can be turned into an
integral over three copies of  $\Sigma$, by trading the three instanton
moduli for the locations of the three insertion points \refs{\DSWW,\Exact}
$$W'^{(k=1)}_{mnp}=\int_C b^{(m)}\int_C b^{(n)}\int_C
b^{(p)}\ e^{-\int_C\omega}$$
Note that, as required, this is holomorphic in the \Ka\ class $\omega$.

For higher $k$, things are quite a bit more complicated. There are more fermi
zero modes, which means we need to bring down powers of the 4-fermi term
from the action. At the same time, the instanton moduli space is
higher-dimensional, and we obtain an, in principle very nontrivial, form that
we have to integrate over it.

First let's count zero modes. After the operator insertion have done their
work, we still have $2k-2$ zero modes of $\lambda^3$ and $k-1$ zero modes
each of $\lambda_1,\lambda_2$ (and the same number of $\psi$ zero modes)
which must be absorbed by bringing down
factors of $S_4$ from the action. In the geometry we are looking at, the only
nonzero components of the Riemann tensor that can absorb the relevant zero
modes are $\int
R^{1\bar1}{}_{3\bar3}\lambda_1\lambda^3\psi_{\bar 1}\psi^{\bar 3}$ and
$\int
R^{2\bar2}{}_{3\bar3}\lambda_2\lambda^3\psi_{\bar 2}\psi^{\bar
3}$. We need to bring down $(k-1)$ factors of each from the action to
absorb the fermion zero modes.

The insight of Aspinwall and Morrison \AspMor\ was that the transformation
properties of the fermion zero modes as a function of the instanton moduli
are simpler in the topological version
of this \sm\ \AandB, leading to a more tractable calculation. This observation
would not be of much use, but for the fact that one can argue that the
3-point
function in the ``A" model $$\tilde W_{mnp}=\vev{{\bi b}^{(m)}(0) {\bi
b}^{(n)}(1){\bi b}^{(p)}(-1)}$$ is equal\foot{for {\it this} choice of
holomorphic 3-form. More generally, they are sections of {\it different}
bundles over the moduli space and are  related by $W'_{mnp}=s^3\tilde
s^{-3}\tilde W_{mnp}$.}\ to the desired $W'_{mnp}$.
This is a triviality for the tree-level result \Wclass; it is far from obvious
that it is true for the instanton corrections which are sensitive to the global
features of the theory. Still, since these 3-point functions {\it are} equal
at the conformal point, it makes sense to calculate them in the topological
theory, where the description of the fermion zero modes is simpler.

The difference between the topological theory and our original \sm\ is that
whereas before all of the fermions were spinors on the world-sheet, now
$\lambda^i,\psi^\ib$ are scalars, and $\lambda_i,\psi_{\ib}$ are 1-forms.
This changes the counting of zero modes. In the topological theory,
 there are   $2k+1$
normalizable zero modes of $\lambda^3,\psi^{\bar 3}$ from \comps{a} and $k-1$
zero modes each of $\lambda_1,\lambda_2,\psi_{\bar 1},\psi_{\bar 2}$.
The operators in the correlation function now absorb 3 zero modes each
of $\lambda^3,
\psi^{\bar 3}$ and we still have to bring down $(k-1)$ factors of
$\int
R^{1\bar1}{}_{3\bar3}\lambda_1\lambda^3\psi_{\bar 1}\psi^{\bar 3}$ and $(k-1)$
factors of $\int
R^{2\bar2}{}_{3\bar3}\lambda_2\lambda^3\psi_{\bar 2}\psi^{\bar
3}$ to absorb the rest.

The simplification that now arises is that the $\lambda^3$ zero modes are
related by supersymmetry to the bosonic zero modes, that is to say, they
transform as tangent vectors to the instanton moduli space\foot{This is the
part that is awkward in the untwisted theory. There the supersymmetry which
relates the bosonic zero modes to zero modes of $\lambda^3$ has a kernel. So
the $\lambda^3$ zero modes transform as sections of a certain subbundle
of $T_\cp{2k+1}$. This complicates the rest of the
argument.}, \ie\ as sections of $T_{\cp{2k+1}}$. An explicit examination
of the zero modes of $\lambda_1,\lambda_2$ show that they transform as sections
of $\CL$, the
tautological line bundle on $\cp{2k+1}$.  The zero modes of $\psi$
transform as the complex conjugates of the $\lambda$'s.

Integrating over the fermion zero modes turns
$\int
R^{1\bar1}{}_{3\bar3}\lambda_1\lambda^3\psi_{\bar 1}\psi^{\bar 3}$ into a
2-form
on $\cp{2k+1}$. It is clear that this 2-form in nothing other than $c_1(\CL)$,
the first Chern class of $\CL$. So
$$\tilde W_{mnp}=\int_{\cp{2k+1}} b^{(m)}\wedge b^{(n)}\wedge b^{(p)}
\wedge c_1(\CL)^{2k-2}\ e^{-S_{inst}}$$
Using the fact that $c_1(\CL)=-H$, where $H$ is the generator of integral
cohomology on $\cp{2k+1}$, this gives
$$\tilde W_{mnp}=\int_C b^{(m)}\int_C b^{(n)}\int_C b^{(p)}\
e^{-k\int_C\omega}$$
which, except for the fact that the instanton action is $k$ times as big, is
exactly the result for $k=1$.

Summing over $k$, we get
$$W'_{mnp}=\int_M b^{(m)}\wedge b^{(n)}\wedge b^{(p)} +\sum_C \sum_{k=0}^\infty
\int_C b^{(m)}\int_C
b^{(n)}\int_C b^{(p)}\ e^{-k\int_C\omega}$$
Clearly, one can sum this geometric series and obtain
$$
W'_{mnp}=\int_M b^{(m)}\wedge b^{(n)}\wedge b^{(p)}+\sum_C \int_C
b^{(m)}\int_C
b^{(n)}\int_C b^{(p)}\ {e^{-\int_C\omega}\over1-e^{-\int_C\omega}}$$
In the simplest case of a 1-dimensional \Ka\ moduli space, we can let
$\omega=t\alpha$,  and simply take $b=\alpha$, where $\alpha$ is a generator
of $\H{2}{M,\BZ}$. Then $\int_C\alpha=n\in\BZ^+$, where $n$ is called the
{\it degree} of the curve $C$. Then
$$W'=\int\alpha^3+ \sum_{n=1}^\infty {a_nn^3e^{-tn}\over 1-e^{-tn}}$$
where $a_n$ is the number of curves of degree $n$. Mirror symmetry
\GPmirror, which
gives one an independent way of calculating $W'$ as the $W$ of some other
Calabi-Yau manifold gives one predictions \CandMir\ for the $a_n$'s. This is
the subject of Candelas's lectures.

\bigbreak\bigskip\bigskip\centerline{{\bf Acknowledgments}}\nobreak
\frenchspacing{
I would like to thank B. Greene and D. Morrison for very helpful discussions
and for their comments on a draft of this manuscript.
This work was supported by
NSF grant PHY90-21984. }

\listrefs
\end